\documentstyle[festkoer,psfig]{vieweg}
\author{Holger Fehske, Michael Holicki, Alexander Wei{\ss}e}
\Kurzautor{Fehske et al.}
\title{Lattice dynamical effects on the Peierls transition  
in one-dimensional metals and spin chains}
\Kurztitel{Lattice dynamical effects on the Peierls transition}
\Adresse{Physikalisches Institut, Universit\"at Bayreuth, D-95440 Bayreuth}
\begin{document}
\def\gsim{\hbox{$\lower1pt\hbox{$>$}\above-1pt\raise1pt\hbox{$\sim$}$}}
\def\lsim{\hbox{$\lower1pt\hbox{$<$}\above-1pt\raise1pt\hbox{$\sim$}$}}
\Titel
\begin{abstract}
The interplay of charge, spin and lattice degrees of freedom
is studied for quasi-one-dimensional electron and spin systems 
coupled to quantum phonons. Special emphasis is put on the influence
of the lattice dynamics on the Peierls transition.  
Using exact diagonalization techniques the ground-state 
and spectral properties of the Holstein model of spinless fermions and 
of a frustrated Heisenberg model with magneto-elastic coupling
are analyzed on finite chains. In the non-adiabatic regime 
a ($T=0$) quantum phase transition from a gapless 
Luttinger-liquid/spin-fluid state to a gapped dimerized 
phase occurs at a nonzero critical value of the electron/spin-phonon 
interaction. To study the nature of the spin-Peierls transition
at finite temperatures for the infinite system,
an alternative Green's function approach is applied to 
the magnetostrictive XY model. With increasing phonon frequency
the structure factor shows a remarkable crossover from soft-mode to 
central-peak behaviour. The results are discussed in relation 
to recent experiments on $\rm CuGeO_3$.
\end{abstract}
\section{Introduction}
\begin{figure}{}
\hfill\psfig{figure=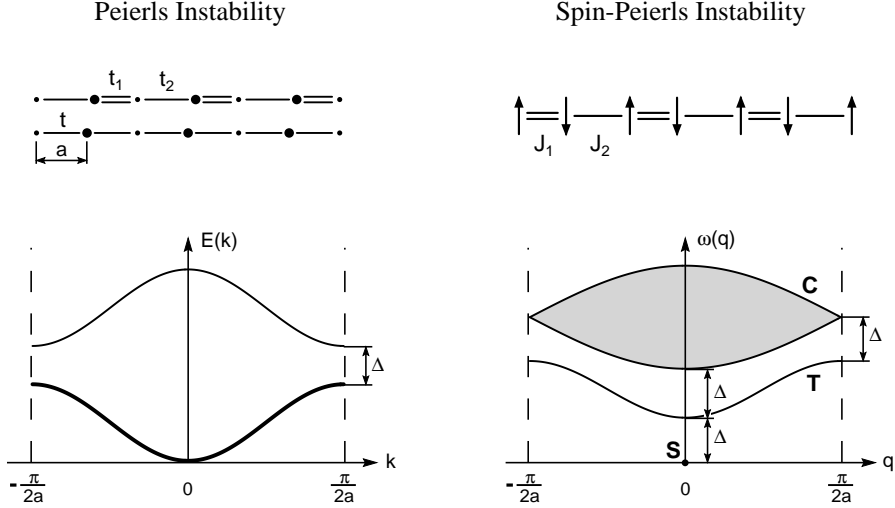,width=12.0cm}\hfill
\caption[fig1]{Schematic representation of the Peierls- and spin-Peierls 
scenarios.  The left panel shows the opening of a gap $\Delta$ 
in the electronic band structure $E(k)$ of an 1D metal at the Fermi 
surface if, according to an SSH- or Holstein-type of electron-phonon 
coupling, a static lattice distortion with the new lattice period $2a$
occurs. The right panel represents the main features of the excitation
spectrum of a Peierls-distorted 1D quantum spin chain. Above the singlet (S)  
ground state at least one elementary excitation, corresponding e.g. to a
triplet (T) bound state, is split from the continuum (C).}  
\label{f1}
\end{figure}

Low dimensional electronic materials are known to be very susceptible
to structural distortions driven by the electron-phonon interaction.
Probably the most famous one is the Peierls instability~\cite{Pe55} of 
one-dimensional (1D) metals: as the temperature is lowered     
the system creates a periodic variation in charge density,    
called a ``charge-density-wave'' (CDW), by shifting
the electrons and ions from their symmetric positions. 
For the half-filled band case the CDW is commensurate with the lattice
and cannot slide as a whole. As a result the unit cell is doubled and  
the system has a broken-symmetry ground state. Since the dimerization 
of the lattice opens a gap at the Fermi surface the Peierls process 
transforms a metal into an insulator. Spontaneous dimerization 
transitions to a less symmetric but lower-energy configuration 
like those shown in Fig.~\ref{f1} (left panel)
have been found in many quasi-1D materials, such as 
the organic conjugated polymers [e.g., (CH)$_x$] and 
charge transfer salts [e.g., TTF(TCNQ)] or the inorganic blue bronzes 
[e.g., $\rm K_{0.3}MoO_3$] and MX-chains~\cite{exsum}.  

As a generic theoretical model for such systems the 1D Hubbard
model is frequently considered,  
\begin{equation}
 {\cal H}_{\rm e} = -\sum_{i,\sigma}t_{ii} n^{}_{i\sigma}
-\sum_{i,\sigma}t_{ii+1} (c^{\dagger}_{i\sigma}c^{}_{i+1 \sigma}
   +\mbox{H.c.}) +U\sum_i n_{i\uparrow}n_{i\downarrow}\,,
\label{humo}
\end{equation}
supplemented by a coupling to the phonon system
\begin{equation}
{\cal H}_{\rm p}=\sum_i\frac{p_i^2}{2M}+\frac{K}{2}q_i^2
\label{phmo}
\end{equation}
according to
\begin{equation}
\begin{array}{r@{\qquad}c@{\qquad}c@{\qquad}c@{\qquad}c}
\mbox{SSH-type~\protect\cite{SSH79}:}
&t_{ii}=0& t_{ii+1} \to t(1+\lambda q_i) & q_i=u_i-u_{i+1}\\
\mbox{Holstein-type~\protect\cite{Ho59a}:}&
 t_{ii}\to \lambda q_i&  t_{ii+1}=t & q_i=u_i\,,
\end{array}
\label{ersetzung}
\end{equation}
i.e., the lattice vibrations $q_i$ interact with the electrons 
by modifying the electron hopping matrix element $t_{ij}$
and on--site potential $t_{ii}$, respectively.
In Eq.~(\ref{humo}), $c^{\dagger}_{i\sigma}$ ($c^{}_{i\sigma}$)
creates (annihilates) a spin-$\sigma$ electron at Wannier site $i$, 
and $n_{i\sigma}=c^{\dagger}_{i\sigma} c^{}_{i\sigma}$.
 
If there is one electron per site and the Coulomb repulsion
is strong, $U\gg t$, we are in the limit of localized electrons
interacting via an effective antiferromagnetic (AF) exchange interaction $J$
($\propto  t^2/U$) and the system can be described by an Heisenberg Hamiltonian
with magneto-elastic coupling~\cite{KGM96}
\begin{equation}
{\cal H} = -J\sum_{i}(1+\lambda q_i) \vec{S}_i\vec{S}_{i+1}
+{\cal H}_{\rm p}\,,
\label{spimo}
\end{equation}
where $\vec{S}_i=\sum_{\sigma\sigma'}
\tilde{c}^{\dagger}_{i\sigma}\vec{\tau}_{\sigma\sigma'}
\tilde{c}^{}_{i\sigma'}$ with $\tilde{c}^{(\dagger)}_{i\sigma}=
c^{(\dagger)}_{i\sigma}(1-\tilde{n}^{}_{i,-\sigma})$. 
In analogy to the Peierls instability in 1D metals,
the dependence of $J$ on the distance 
between neighbouring spins again gives rise to an instability, 
where the energy of the spin chain is lowered by 
dimerizing into an alternating pattern of weak and strong  
bonds. This so-called ``spin-Peierls'' (SP) transition leads to
the formation of a singlet (dimer) ground state and there
is an energy gap to elementary (massive) spin triplet excitations
being well-separated from the continuum [see Fig.~\ref{f1} (right
panel)]. Experimentally, the SP phenomenon was observed in a number
of organic  compounds, such as (TTF)(CuBDT) or MEM(TCNQ)$_2$~\cite{BIJB83}.

Most theoretical treatments of the Peierls instability describe
the lattice degrees of freedom classically. 
In a wide range quasi-1D metals, however, the lattice zero-point motion 
is comparable to the Peierls lattice distortion, which makes 
the rigid lattice approximation questionable~\cite{MW92}. 
By any means lattice dynamical (quantum phonon) effects should be included 
in a theoretical analysis of the extraordinary transport and optical 
phenomena observed in Peierls-distorted systems~\cite{HS85,Deea95}.
Likewise the interest in models of spins coupled dynamically  
to phonons has increased significantly since it was recognized 
that the first inorganic SP compound $\rm CuGeO_3$~\cite{HTU93} shows no 
clear separation between the magnetic and phononic energy scales:
the two (weakly dispersive optical) Peierls-active 
$\rm T_2^+$ phonon modes have frequencies $\omega_{0,1}\simeq J$ and 
$\omega_{0,2}\simeq 2J$~\cite{Brea96,WGB99}.  
Moreover, in contrast to the organic SP materials, no phonon-softening 
is observed at the SP transition~\cite{Loea94}. 
The SP physics in $\rm CuGeO_3$ is therefore in the non-adiabatic regime. 

Motivated by this situation, in this report, we study the perhaps   
minimal microscopic models capable of describing the Peierls and 
spin-Peierls transitions in 1D systems by the use of numerical 
techniques allowing an essentially exact treatment of both 
subsystems, electrons/spins and phonons, 
at a fully quantum-mechanically level.   
In the weak-coupling regime, the random-phase-approximation
(RPA) approach is shown to be consistent not only with phonon
softening but also with phonon hardening at the SP transition,
as observed, e.g., in (TTF)(CuBDT) and $\rm CuGeO_3$, respectively.
\section{Luttinger-liquid vs. charge-density-wave behaviour}
First we consider the 1D Hubbard model, Eq.~(\ref{humo}), at quarter filling 
and confine ourselves to the case of spinless fermions for simplicity.
This model is of physical relevance in the strong interaction limit
$U\to\infty$ and is particularly interesting because a quantum phase 
transition from a Luttinger liquid (LL) to a CDW phase occurs at a finite 
electron-phonon interaction, demonstrating that the quantum phonon 
fluctuations destroy the dimerized ground state for weak electron-phonon 
couplings~\cite{HF83,BMH98}.

Rescaling ${\cal H}\to{\cal H}/t=\bar{\cal H}_{\rm e}
+\bar{\cal H}_{\rm p} +\bar{\cal H}_{\rm ep}$ and setting 
\begin{equation}
\lambda=g\sqrt{2K\omega_0}\qquad\mbox{with}\qquad \omega_0^2=K/M
\label{laom}
\end{equation}
the Holstein Hamiltonian in a particle-hole symmetric notation  is given by 
\begin{equation}
\label{homosf}
\bar{\cal H}_{\rm e} = 
- \sum_i ( c_i^{\dagger} c_{i+1}^{} + c_{i+1}^{\dagger} c_{i}^{} )\;,\;\;\;
\bar{\cal H}_{\rm p} =  \omega_0 \sum_i  ( b_i^{\dagger} b_i^{}+ \mbox{\small 
$\frac{1}{2}$})\,,
\end{equation}
and
\begin{equation}
\bar{\cal H}_{\rm e-p}= - g \omega_0 \sum_i ( b_i^{\dagger}  
+ b_i^{})\,  (n_i^{} 
-\mbox{\small $\frac{1}{2}$})\;.
\label{sfmo}
\end{equation}
In Eqs.~(\ref{homosf}) and (\ref{sfmo}),  
$b^{(\dagger)}_i$ annihilates (creates) a dispersionsless 
Einstein phonon of frequency $\omega_0$ coupled to the local 
electron density $n_i=c^{\dagger}_ic_i^{}$
($\hbar=1$, and all energies are measured in units of $t$).
Note that for the Holstein model the dimensionless electron-phonon 
coupling constant $g=\sqrt{\varepsilon_p/\omega_0}$ is directly
related to the familiar polaron shift $\varepsilon_p$ being 
a second natural measure of the strengths of the 
electron-phonon interaction. Both parameters are necessary
in order to characterize the weak ($\varepsilon_p \ll 1$) and 
strong coupling ($\varepsilon_p \gg 1$ {\it and} $g \gg 1$)
situations in the adiabatic ($\omega_0\ll 1$) and 
anti-adiabatic regimes ($\omega_0\gg 1$), respectively.

\begin{figure}
\hfill\psfig{figure=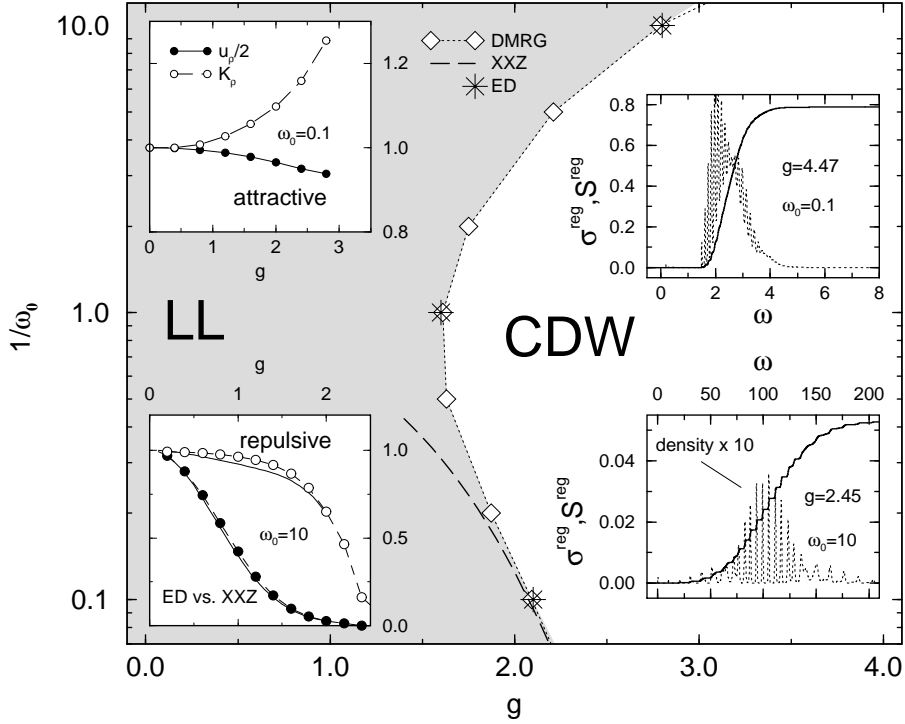,width=12.0cm}\hfill
\caption[fig2]{Ground-state phase diagram of the 1D
Holstein model of spinless fermions at half-filling ($N_e=N/2$),
showing the boundary between the Luttinger liquid (LL)  
and charge-density-wave (CDW) states obtained by 
exact diagonalization (ED) and density matrix renormalization 
group (DMRG)~\cite{BMH98} approaches. The dashed line gives the asymptotic
result for the XXZ model. Left insets show the 
LL parameters  $u_{\rho}$ and $K_{\rho}$ as a function of the 
electron-phonon $g$ in the metallic regime; right
insets display for a six-site chain the regular part of the optical 
conductivity $\sigma^{reg}(\omega)$ (dotted lines) and the integrated spectral 
weight ${\cal S}^{reg}(\omega)=\int_0^{\omega}d\omega^{\prime}
\sigma^{reg}(\omega^{\prime})$ (solid lines) in the CDW region.}    
\label{f2}
\end{figure}

Previous results for the ground-state
phase diagram of the Holstein model at half-filling 
obtained by WL QMC~\cite{HF83} and GF MC~\cite{MHM96} simulations
showed significant discrepancies in the region of small $\omega_0$ 
($0 < \omega_0 \lsim 1$). Only very recently Bursill 
et al.~\cite{BMH98} provided more reliable 
information from level crossings in their DMRG data.
Applying a new optimized phonon approach for the diagonalization 
of coupled electron/spin-phonon
systems~\cite{WFWB00}, we consider the Holstein model
on chains of even length with up to 10 sites and periodic (antiperiodic) 
boundary conditions if there is an odd (even) number of fermions in the
system. The resulting phase diagram is shown in Fig.~\ref{f2}
over a wide range of frequencies and coupling strengths. 
For small $g$ the system is a metal, more precisely a Luttinger
liquid with parameters that vary with the coupling (see below).
For large $g$ the system has an energy gap and develops true 
long-range CDW order in the thermodynamic limit. 
The phase boundary obtained with our optimized phonon diagonalization
methods agrees with recent DMRG results~\cite{BMH98}. In the adiabatic limit
$\omega_0=0$, the critical coupling converges to zero, as expected 
for the Holstein Hamiltonian, Eqs.~(\ref{humo})-(\ref{ersetzung}), 
with $M\to\infty$. For $0< \omega_0 \lsim 1$  
the accurate determination of $g_c$ is somewhat difficult.
On the other hand, in the strong-coupling anti-adiabatic regime, 
the half-filled Holstein model can be transformed to the exactly 
soluble XXZ (small polaron) model~\cite{HF83,YY66}
\begin{equation}
\label{xxz}  
\bar{\cal H}^{\rm XXZ} = 
{N\over 4}(2\omega_0-g^2\omega_0 -V_2) -\mbox{e}^{-g^2} \sum_i\Big[
          (S_i^+ S_{i+1}^- + S_i^- S_{i+1}^+) -
          V_2 \,\mbox{e}^{g^2}  S_i^z S_{i+1}^z \Big]
\end{equation}
using second order perturbation theory with respect to $t$. 
Here $V_2(g^2,\omega_0)= 2 e^{-2g^2} \omega_0^{-1}\sum_{s\ne0} 
(2g^2)^s/(s s!)$, and the (Kosterlitz-Thouless) phase transition 
line is given by the condition 
$V_2(\alpha,g^2)\mbox{e}^{g^2}/2=1$ (long-dashed curve in Fig.~\ref{f2}).

Let us now characterize the LL and CDW phases in some more detail.
According to Haldane's Luttinger liquid conjecture~\cite{Ha80}, 1D 
gapless systems of interacting fermions should belong
to the same universality class as the Tomonaga-Luttinger model.
As stated above, the Holstein system is gapless for small 
enough coupling $g$. Thus it is obvious to prove, 
following the lines of approach to the problem by 
McKenzie~et~al.~\cite{MHM96}, whether our Lanczos data 
shows a finite-size scaling like a Luttinger liquid. 
For a LL of spinless fermions, the ground-state 
energy $E_0(N)$ of a finite system of $N$ sites 
scales to leading order as 
\begin{equation}
\label{uro}
  {E_0(N)\over N} = \epsilon_{\infty} - {\pi u_{\rho}\over 6 N^2} \,,
\end{equation}
where $\epsilon_{\infty}$ denotes the ground-state energy per site for
the infinite system and $u_{\rho}$ is the velocity of the charge excitations.
If $E_{\pm 1}(N)$ is the ground-state energy with $\pm 1$ fermions
away from half filling, to leading order the scaling should be
\begin{equation}
\label{kro}
  E_{\pm 1}(N) - E_0(N) = {\pi u_{\rho} \over 2 K_{\rho} N}\,.
\end{equation}
$K_{\rho}$ is  the renormalized effective coupling (stiffness)
constant. 
The left insets of Fig.~\ref{f2} show the LL parameters 
as a function of $g$ in the region, where the scaling relations 
Eqs.~(\ref{uro}) and (\ref{kro}) hold. 
A very interesting result is the changing character 
of the interaction below $\omega_0\sim 1$. 
For small frequencies the effective fermion-fermion interaction
is attractive, while it is repulsive for large frequencies,
where the system forms a polaronic metal with strongly reduced 
kinetic energy~\cite{WFWB00}.  Self-evidently there is a transition line in  
between, where the model describes ``free'' particles 
in lowest order. 

In the CDW state extremely valuable information 
about the low-energy excitations can be obtained
from the behaviour of the optical conductivity.
The real part of $\sigma(\omega)$ contains two contributions, the 
(coherent) Drude part at $\omega=0$ and a so-called ``regular term'',
$\sigma^{reg}(\omega)$, due to finite-frequency dissipative optical
transitions to excited quasiparticle states. In spectral representation 
($T=0$), the regular part takes the form 
\begin{equation}
\label{sigreg} 
\sigma^{reg}(\omega)=\sum_{m > 0}
\frac{|\langle {\mit \Psi}_0^{} |i \sum_{j}( c_{j}^{\dagger}
 c_{j+1}^{} - c_{j+1}^{\dagger}c_{j}^{}) |  {\mit \Psi}_m^{} 
       \rangle |^2}{E_m-E_0} \;\delta[\omega -(E_m-E_0)]\,,
\end{equation}
where $\sigma^{reg}(\omega)$ is given in units 
of $\pi e^2$ and we have omitted an $1/N$ prefactor. 
The evaluation of dynamical correlation functions, 
such as Eq.~(\ref{sigreg}), can be carried out 
by means of very efficient and numerically stable Chebyshev recursion and 
maximum entropy algorithms~\cite{BWF98}. 
Clearly the optical absorption spectrum in the strong EP coupling regime
is quite different from that in the LL phase (cf. Ref.~\cite{WF98a}).
It can be interpreted in terms of strong electron-phonon correlations 
and corroborates the CDW picture. Since for $g>g_c$ the electronic 
band structure is gapped we expect that the low-energy gap feature
survives in the thermodynamic limit. In the adiabatic region 
(upper right inset), the broad optical absorption band is 
produced by a single-particle excitation accompanied 
by multi-phonon absorptions and is basically related to 
the lowest unoccupied state of the upper band of the CDW insulator. 
The lineshape of~$\sigma^{reg}(\omega)$  
reflects the phonon distribution in the ground state.  
The most striking feature is the large spectral weight
contained in the incoherent part of optical conductivity.
Moreover, employing the f-sum rule for the optical 
conductivity~\cite{WF98b} and taking into account 
the behaviour of the kinetic energy $(\propto u_{\rho})$ 
as function of $g$, we found that in the metallic LL and 
insulating CDW phases nearly all the spectral weight is contained in the 
coherent (Drude) and incoherent (regular) part 
of $\mbox{Re}\ \sigma(\omega)$, respectively.
As stated above, in the anti-adiabatic regime the LL phase is basically a
polaronic metal, i.e., the electrons will be heavily 
dressed by phonons. Since the renormalized coherent 
bandwidth of the polaron band is extremely small, the finite-size gaps in 
the band structure are reduced as well. Therefore, the gap occurring  
in the CDW state ($\Delta_{CDW}\sim \varepsilon_p$)   
may be identified with the optical absorption threshold 
(see lower right inset).
\section{Non-adiabatic approach to the spin-Peierls transition}
\subsection{Exact diagonalization results for T=0}
In spite of the experimental fact that a realistic modeling of the 
inorganic SP compound $\rm CuGeO_3$ should include 
the phonon dynamics, previous theoretical studies 
have commonly adopted an alternating and 
frustrated AF Heisenberg spin chain 
model~\cite{CCE95}
\begin{equation}
\bar{\cal H}^{\rm static}=
\sum_{i}\left[(1+\delta(-1)^i) \vec{S}_{i}\vec{S}_{i+1} \,+\alpha\,
 \vec{S}_{i}\vec{S}_{i+2} \right]
\label{sfhm}
\end{equation}
with a static dimerization parameter $\delta$, thus representing the extreme 
adiabatic limit of a SP chain (in this section all energies are given
in units of $J$). $\alpha$ determines the strength of the frustrating 
AF next-nearest-neighbour coupling. The spin model~(\ref{sfhm})   
contains two independent mechanisms for spin gap 
formation. At $\delta=0$ and for $\alpha<\alpha_c$ the ground 
state is a spin liquid and the elementary excitations are
massless spinons \cite{Ha82}. The critical value of frustration 
$\alpha_c=0.241$ was accurately determined by numerical studies 
\cite{CCE95,ON93}. For $\alpha>\alpha_c$ the ground state is spontaneously
dimerized, the spectrum acquires a gap, and the elementary excitations are 
massive spinons~\cite{CPKSR95,WA96}. On the other hand 
for any finite $\delta$, the singlet ground state of the model~(\ref{sfhm}) 
is also dimerized, but the elementary excitation is a massive magnon 
\cite{Ha82,Ts92}. A comprehensive study of the spectral properties of
the model~(\ref{sfhm}) in terms of the spin dynamical structure factor
has been carried out by Yokoyama and Saiga~\cite{YS97}.

From the magnetic properties of the uniform phase $J\simeq 160$~K and 
$\alpha=0.36$ have been estimated for $\rm CuGeO_3$~\cite{Faea97}. 
However, if one attempts to reproduce the observed spin gap 
$\Delta^{\rm ST}\simeq 2.1~\rm{meV}$ within the static model~(\ref{sfhm}), 
a very small value of $\delta\simeq 1.2\%$ results. 
From the uniaxial pressure derivatives of the exchange coupling
$J$~\cite{BFKW99} and the structural distortion in the dimerized 
phase~\cite{Brea96} a minimum magnetic dimerization of about 4\% 
is obtained, incompatible with an adiabatic approach to the SP transition. 

\begin{figure}
\hfill\psfig{figure=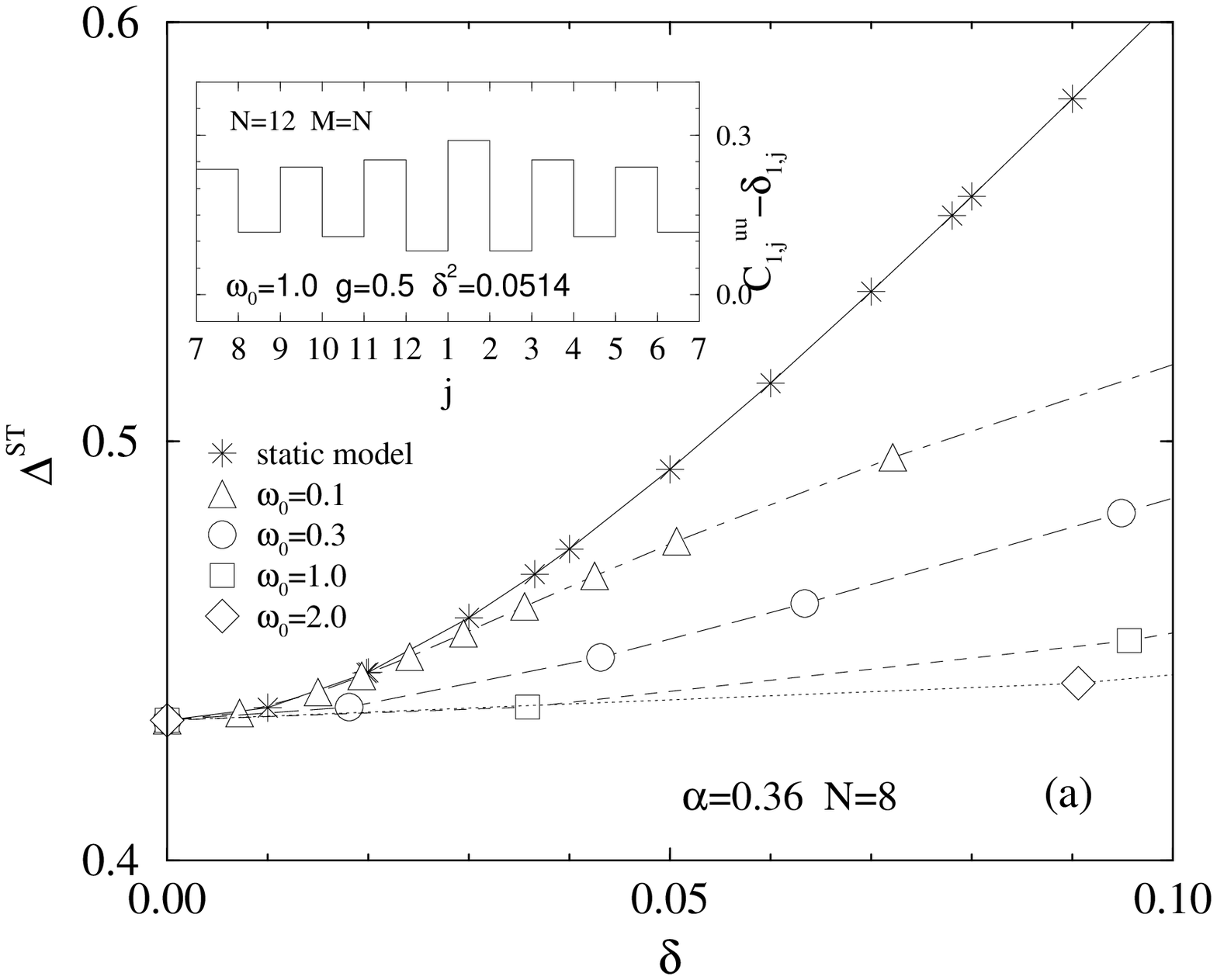 ,width=6.0cm}
\hfill\psfig{figure=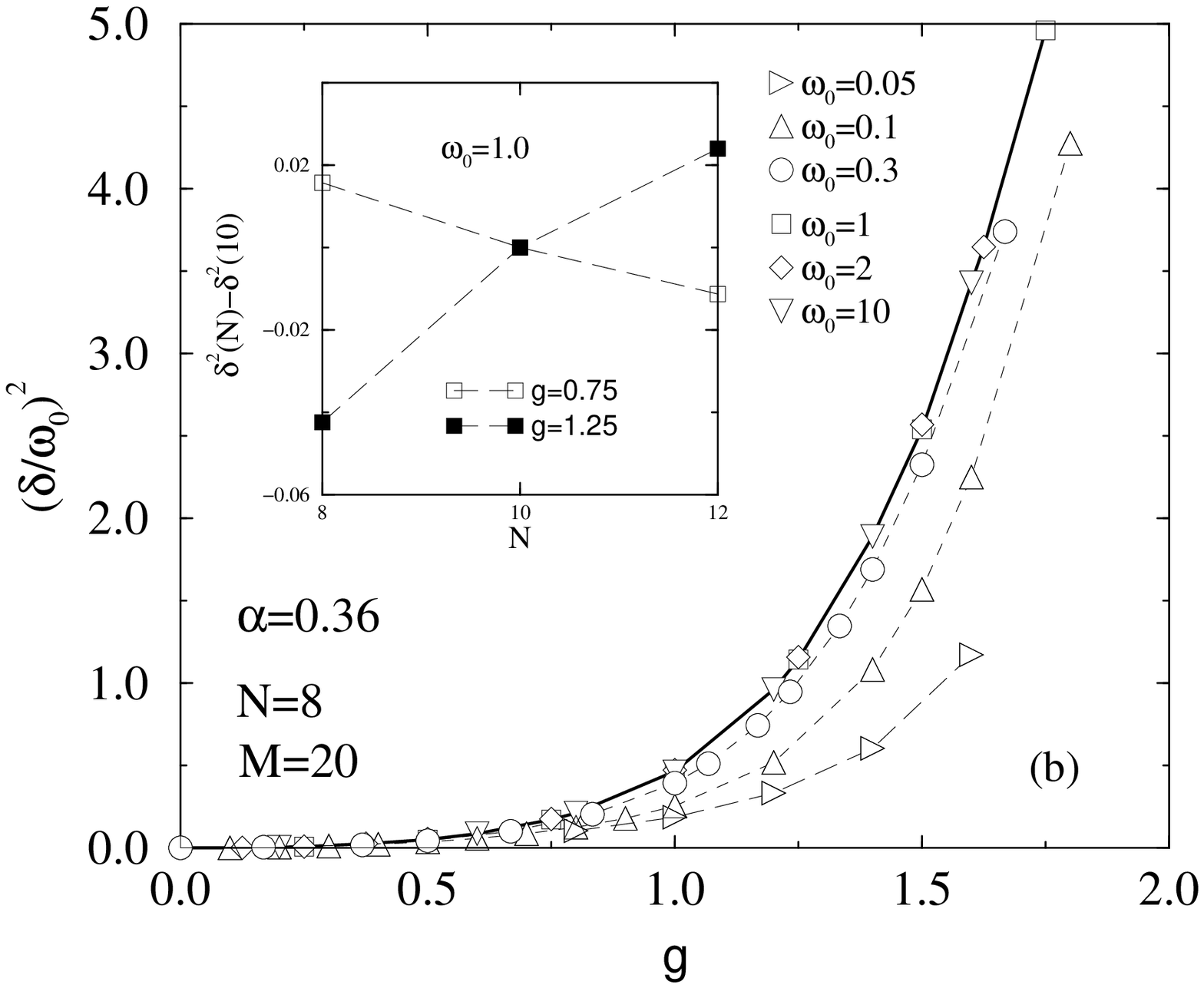,width=6.0cm}\hfill
\caption[fig3]{Spin gap (a) and dimerization (b) in the frustrated Heisenberg
spin chain with dynamical spin-phonon coupling.}
\label{f3}
\end{figure}

The simplest model that maintains the full quantum dynamics of the 
lattice vibrations may be obtained from Eq.~(\ref{sfhm}) by replacing
$(-1)^i\delta \to g\omega_0(b_i^{\dagger}+b_i^{})$: 
\begin{equation}
\bar{\cal H}_{\rm s} =  \sum_i (\vec{S}_{i}\vec{S}_{i+1} 
+ \alpha\vec{S}_{i}\vec{S}_{i+2}  )\,,\;\;\;
\bar{\cal H}_{\rm p} =  \omega_0 \sum_i b_i^{\dagger} b_i\,,
\label{fhmo}
\end{equation}
\begin{equation}
 \bar{\cal H}_{\rm sp}^{\rm I} 
= g \omega_0 \sum_i (b_i^{\dagger} + b_i)\vec{S}_{i}\vec{S}_{i+1}\,.
\label{spcI}
\end{equation}
Recently it was shown that such a dynamical spin-phonon model 
describes the general features of the magnetic excitation spectrum
of $\rm CuGeO_3$~\cite{WFK98,APSA98,SC99}. 
Here we focus on the behaviour of the lattice dimerization which can
be found from the displacement structure factor at wave number $q=\pi$ 
\begin{equation}
\delta^2=\left(\frac{g\omega_0}{N}\right)^2\sum_{i,j}C_{ij}^{uu}
\mbox{e}^{{\rm i}\pi(R_i-R_j)}\quad\mbox{with}\quad
 C_{ij}^{uu}=\langle(b_i^{}+b_i^{\dagger})(b_j^{}+b_j^{\dagger})\rangle\;.
\label{d2}
\end{equation} 
The alternating structure of the correlation function $C_{1i}$ as shown in 
the inset Fig.~\ref{f3}~(a) implies the Peierls formation 
of short and long bonds and thus alternating 
strong and weak AF exchange interactions, i.e., a dimerized ground state. 
The structure is enhanced (weakened) increasing the spin--phonon 
coupling (phonon frequency)~\cite{WFK98}.
As in the ordinary Peierls phenomenon, a finite dimerization 
$\delta>0$ necessarily leads to a gap $\Delta^{\rm ST}$ in the 
magnetic excitation spectrum. For evaluating the relation 
between the dimerization and the resulting magnitude of the spin triplet 
excitation gap, we keep the phonon frequency fixed, vary the coupling 
strength $g$ and calculate for each parameter set 
$\Delta^{\rm ST}=E_0^{\rm T}-E_0^{\rm S}$ and the dimerization 
$\delta$ from Eq.~(\ref{d2}).  ED results for $\Delta^{\rm ST}$ 
obtained for the static spin-only model~(\ref{fhmo}) and the 
quantum phonon model~(\ref{spcI})  are compared in the main part of 
Fig.~\ref{f3}~(a). For vanishing dimerization, 
i.e. in the absence of any spin--phonon coupling $(g=0)$ the results 
for the static and the dynamic model naturally agree 
(note that for $\delta=0$ there remains a spin excitation gap 
due to the frustration driven singlet dimer ordering).
The dynamic model~(\ref{spcI}) partially resolves the $\Delta^{\rm ST}-\delta$ 
conflict we are faced within the static approach,
because the dimerization $\delta$ grows with the phonon frequency  
at fixed $\Delta^{\rm ST}$. Thus matching $\Delta^{\rm ST}$ 
to the neutron scattering data, a larger $\delta\simeq$ 5\% may result. 

Figure~\ref{f3}~(b) shows the dimerization as a function of the 
spin-phonon interaction strength. As for the Holstein model
the precise determination of the critical coupling $g_c$ is  
difficult for phonons in the adiabatic regime $\omega_0 < 0.5$. 
On the contrary, in the non-adiabatic regime the SP transition 
takes place at $g_c \simeq 1$ nearly irrespective of $\omega_0$. 
Most remarkably, below (above) $g_c$ the dimerization decreases 
(increases) with increasing lattice size $N$ (see inset), 
indicating that the infinite system exhibits a true phase transition.
\subsection{Anti-adiabatic limit: mapping to an effective magnetic problem}
Figure~\ref{f3}~(b) implies that more insight into the dynamic spin-phonon 
model can be obtained by considering the limit of large phonon frequencies.
One can then integrate out the lattice degrees of freedom in order to
derive an effective spin Hamiltonian which cover the dynamical effect 
of phonons in a approximate way. Technically this can be achieved by 
a variety of methods, such as standard perturbation theory~\cite{KF87}, 
continuous (flow-equation based)~\cite{Uh98} or variational unitary 
transformations~\cite{WWF99}.  

In what follows we consider besides the 
magneto-elastic interaction, Eq.~(\ref{spcI}), 
another type of spin-phonon coupling~\cite{KF87,BMH99}, 
\begin{equation}
\bar{\cal H}_{\rm s-p}^{\rm II} =  g \omega_0  \sum_i (b_i^{\dagger} + b_i)
  (\vec{S}_{i}\vec{S}_{i+1} -\vec{S}_{i}\vec{S}_{i-1} )\,,
\label{spcII}
\end{equation}
where the AF exchange integral varies linearly with the difference between 
the phonon amplitudes on neighbouring sites, and perform a 
Schrieffer-Wolff transformation~\cite{SW66},
$\tilde{\bar{\cal H}}  =  \exp({\cal S}) \bar{\cal H} \exp(-{\cal S})$, 
with 
\begin{equation}
  {\cal S}^{\rm I}  =  g 
  \sum_i (b_i^{\dagger} - b_i) \vec{S}_{i}\vec{S}_{i+1}\quad\mbox{and}\quad  
{\cal S}^{\rm II} =  g  
  \sum_i (b_i^{\dagger} - b_i)(\vec{S}_{i}\vec{S}_{i+1} 
-\vec{S}_{i}\vec{S}_{i-1})\,,
\end{equation}
respectively. In contrast to electron-phonon systems with Holstein coupling, 
where the (Lang-Firsov) transformation similar to $\exp(S)$ completely
removes the electron-phonon interaction term ($\bar{\cal H}_{\rm e-p}$), 
applying the unitary transformation $\exp(S)$ to 
$\bar{\cal H}$ with~Eqs.~(\ref{spcI}) and~(\ref{spcII}), 
we obtain an infinite series of terms, which can 
not be summed up to a simple expression. To derive an effective 
spin model we now take the average over the (transformed) phonon vacuum. 
The resulting spin Hamiltonian  $\bar{\cal H}_{\rm eff,s}=\langle 
\tilde{0}_{\rm p}|\tilde{\bar{\cal H}}|\tilde{0}_{\rm p}\rangle$  
contains longer than next-nearest-neighbour ranged Heisenberg interactions 
as well as numerous multi-spin couplings. 
To a good approximation we can neglect them and obtain~(cf.~\cite{WWF99})
\begin{equation}
\bar{\cal H}_{\rm eff,s}^{\rm I/II} = 
\sum_i (\vec{S}_{i}\vec{S}_{i+1} 
+ \alpha_{\rm eff}^{\rm I/II} \vec{S}_{i}\vec{S}_{i+2}  ) 
\label{efffhmo}
\end{equation}
with
\begin{equation}
\alpha_{\rm eff}^{\rm I}= \frac{\alpha +g^2(1-2\alpha)/2
+3g^4\omega_0/ 16 -37 g^4 (1-2\alpha)/ 96}{ 
1+ g^2\omega_0/2 -g^2(1-\alpha)/2 - 3 g^4\omega_0/ 8
+g^4 (28-37\alpha)/96}\,,
\label{alphaI}
\end{equation}
\begin{equation}
\alpha_{\rm eff}^{\rm II}=\frac{\alpha +g^2\omega_0/2 +g^2(3-5\alpha)/2
+3g^4\omega_0/48-g^4(75-124\alpha)/24}{1+g^2\omega_0-3g^2(1-\alpha)/2-9
g^4\omega_0/8+g^4(59-75\alpha)/24}\,.
\label{alphaII}
\end{equation}
That is the integration over the phonon subsystem   
yields an additional frustrating next-nearest-neighbour 
exchange interaction. Therefore, without any explicit frustration 
$\alpha$, the effective frustration $\alpha_{\rm eff}$ 
due to the phonons can lead to a gap in the 
energy spectrum and to spontaneous dimerization. 
This effect is most important in the anti-adiabatic 
frequency range~\cite{WWF99}.

\begin{figure}
\hfill\psfig{figure=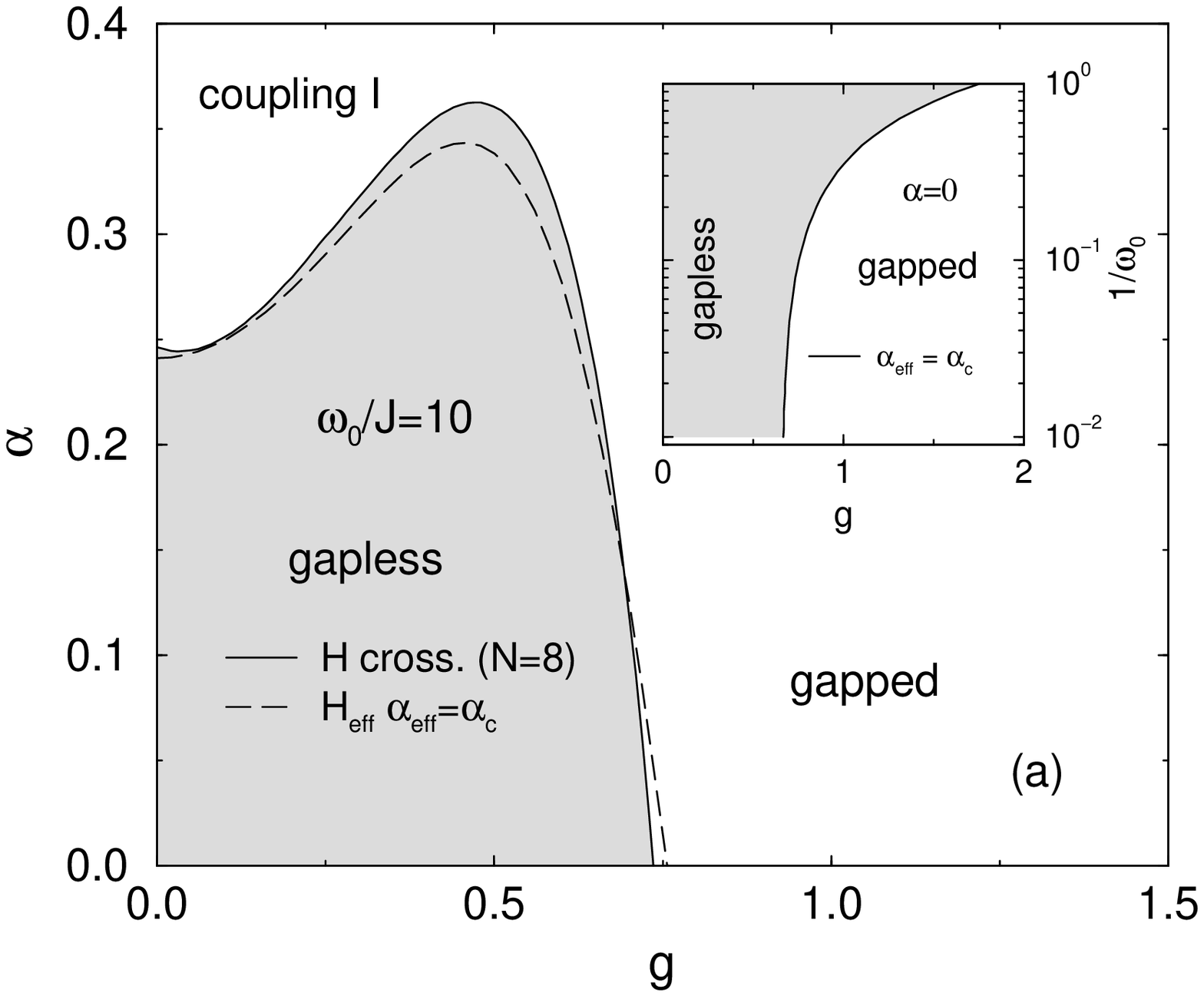,width=6.0cm}
\hfill\psfig{figure=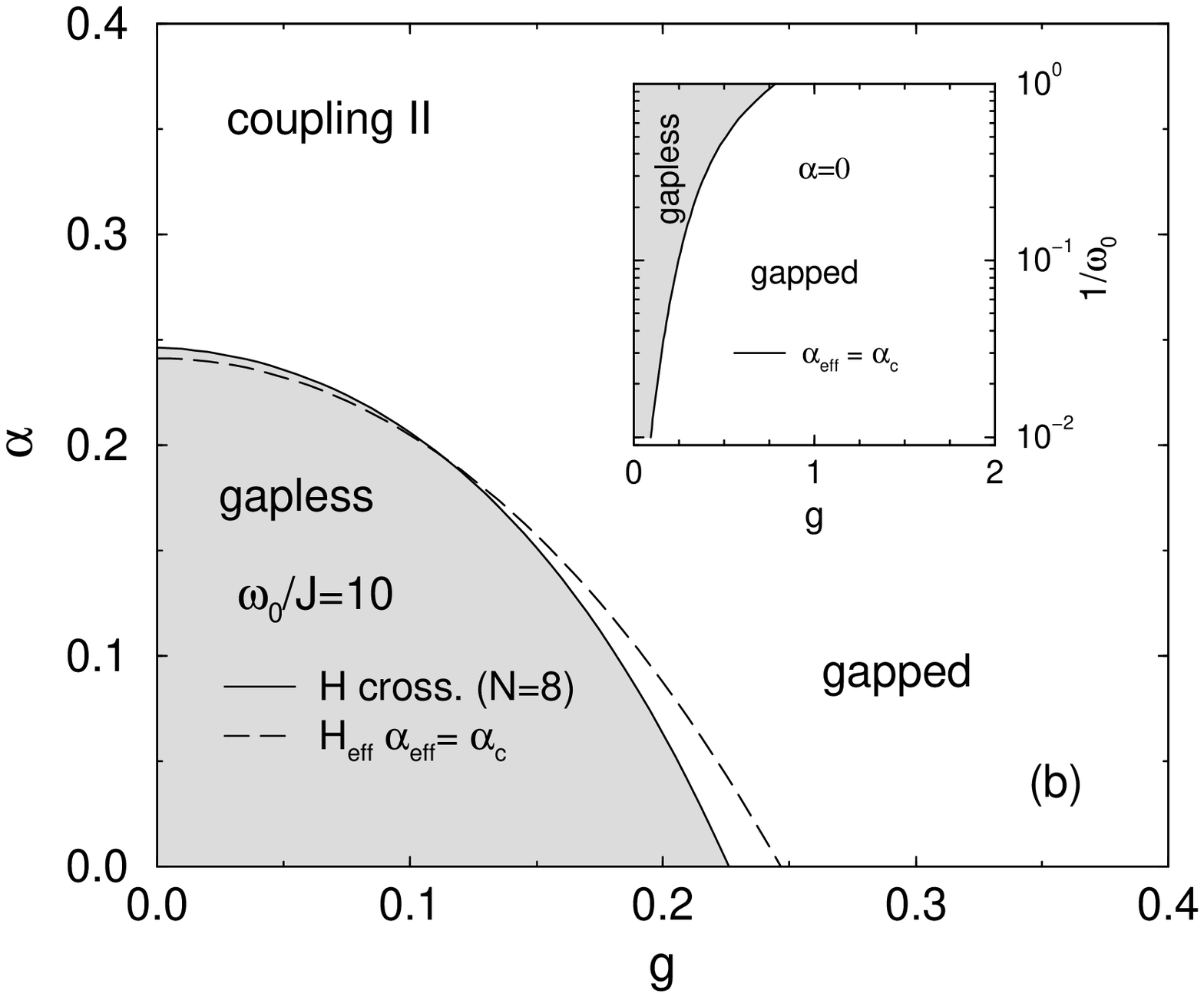,width=6.0cm}
\caption[fig4]{Phase diagram of the original spin-phonon model~(\ref{fhmo})
and the effective spin model~(\ref{efffhmo}) with  
spin-phonon couplings of types~I (a) and~II (b).}
\label{f4}
\end{figure}

In order to determine the critical line in the
the $\alpha-g$ plane  indicating the transition from the 
gapless to the gapped ground state, we use the level
crossing criterion~\cite{ON93,CCE95,WWF99,BMH99}
for the lowest singlet and triplet excitations
which become degenerate at $\alpha_c(N)$. At $\alpha_c$ the 
finite-size corrections $\alpha_c(N)-\alpha_c(\infty)\sim N^{-2}$
are small. The resulting phase diagrams are displayed in 
Fig.~\ref{f4}. We find that the phase boundary of the 
original quantum spin-phonon and effective spin models 
agree surprisingly well, where for the effective model~(\ref{efffhmo}) 
$\alpha_c(g)$ can be obtained with high accuracy on local
workstations if $N\leq 20$. For spin-phonon coupling of type~I 
the critical curve exhibits a remarkable upturn before crossing 
the abscissa; i.e., the frustration is  suppressed for 
small spin-phonon coupling, but over-critical for strong coupling. 
It is this feature which makes 
it necessary to expand $\tilde{\bar{\cal H}}$ up to fourth order in $g$ 
to approximate $\bar{\cal H}$ in a correct way. A second order theory 
is not capable to describe the observed critical line. 
The upturn of $\alpha_c(g)$ at small $g$ was reproduced quite recently 
by linked series expansion techniques~\cite{TEM99}.
By contrast $\alpha_c(g)$  is a monotonous decreasing
function of $g$ for the coupling of type~II. It appears that one 
would get the same shape also for a second order theory. 
However, to enlarge the application area of our approximation taking into 
account higher order contributions is still appropriate.
For $\alpha\equiv 0$, differently from the coupling case~I, 
$g_c$ tends to zero in the anti-adiabatic limit $\omega_0 \to \infty$.
While the $q=0$ and the $q=\pi$ phonon mode compete in the 
case of coupling~I, allowing for a stable
gapless phase up to a critical $g$, there is no interaction
with the $q=0$ mode in $H_{\rm sp}^{\rm II}$. Therefore the 
$q=\pi$ mode induces long ranged exchange more efficiently,
leading to a vanishing $g_c$ for $\omega_0\rightarrow\infty$.
\subsection{ RPA approach at $\bf T\neq 0$: 
Soft-mode vs. central-peak behaviour}
The absence of a soft phonon mode at the displacive SP in $\rm CuGeO_3$
has been puzzling for a long time, since the behaviour was believed to be
inconsistent with the standard Cross-Fisher RPA 
approach to SP transitions~\cite{CF79}. 
Even worse, the relevant phonon modes harden by about 5\% with
decreasing temperature~\cite{Brea98}, pointing to a central peak scenario. 
A way to reconcile these results within the framework
of the Cross-Fisher theory has been proposed by Gros 
and Werner~\cite{GW98}, with the result that a complete phonon
softening occurs only for bare phonon frequencies less than 
a critical value ($\omega_0 < \omega_{0,c}=2.2 T_c$, where $T_c$ is the SP 
transition temperature. For higher $\omega_{0}$ a central peak develops
reaching $T_c$ from above. For the magnetostrictive Heisenberg spin 
chain model, however, it was not possible to analyze the complete
pole structure of phonon spectral function.  

To test this scenario we restrict ourselves to the simpler XY model
\begin{equation}
\bar{\cal H}_{\rm s}^{\rm XY} = \sum_i (S_i^xS_{i+1}^x+S_i^yS_{i+1}^y)\,,
\;\;\; \bar{\cal H}_{\rm p}=\sum_i \frac{p_i^2}{2M}+\frac{K}{2}
(u_i-u_{i+1})^2
\label{xymo}
\end{equation}
with the spin-phonon coupling term 
\begin{equation}
\bar{\cal H}_{\rm s-p}=\frac{\lambda}{2}\sum_i(u_i-u_{i+1})
(S_i^+S_{i+1}^-+S_i^-S_{i+1}^+)\,.
\label{xyspc}
\end{equation}
The above magnetostrictive XY model 
has been proposed as the minimal model to describe the 
SP transition~\cite{CM96}. Performing a Jordan-Wigner
transformation~\cite{JW28} it can be mapped onto a model of spinless 
fermions only interacting with the phonon system
$\bar{\cal H}_{\rm p}=\sum_q\omega_{q}b_q^{\dagger}b_q^{}$:  
\begin{eqnarray}
\label{jwmo}
\bar{\cal H}_{\rm s}^{\rm XY}&\to&
\bar{\cal H}_{\rm e}^{\rm JW} = -\sum_k \cos(k) c_k^{\dagger}c_k^{}\,,\\
\bar{\cal H}_{\rm s-p}&\to&\bar{\cal H}_{\rm e-p}^{\rm JW}=g\omega_0
\sum_{k,q}\eta(k,q)(b_q^{}+b_{-q}^{\dagger})
c^{\dagger}_kc^{}_{k-q}\,,
\end{eqnarray}
where
\begin{equation}
\eta(k,q)=-i\left(\frac{\omega_{0}}{N\omega_{q}}\right)^{\frac{1}{2}}
\left[\sin(k-q)-\sin(k)\right]\,.
\label{eta}
\end{equation}
with  $\omega_{q}=2\omega_{0}\sin(q/2)$, i.e $\omega_{\pi}=2\omega_{0}$.

Determining the equation of motion for the Matsubara phonon Green's function
$\bar{D}(q,i\omega_n)$ within the scheme worked out by 
Bennett and Pytte for the magnetostrictive
Heisenberg model~\cite{BP67}, the main advantage is that the higher-order 
spin-spin correlation functions can be calculated without further 
approximations for the XY-case. The retarded Green's function 
$\bar{D}^{\rm ret}(q,\omega)$ is obtained from
$\bar{D}(q,i\omega_n)$ by analytical continuation  
$i\omega_n \to \omega$. In the uniform phase above $T_c$, 
the RPA propagator of the $q=\pi$-phonon, 
being responsible for the phase transition, then results as  
\begin{equation}
\bar{D}^{\rm ret}(\pi,\omega)=
\frac{4\omega_0}{\omega^2-4\omega_0^2-2\omega_0\bar{P}(\pi,\omega)},
\label{d}
\end{equation}
\begin{equation}
\bar{P}(\pi,\omega)=-(2g\omega_0)^2
\int_{0}^{\pi}\frac{dk}{\pi}
\frac{\sin^2(k) \tanh[\frac{\beta}{2}
\cos(k)]}{\omega+2\cos(k)}.
\label{p}
\end{equation}
A structural instability occurs if $\bar{D}^{\rm ret}(q,\omega)$ exhibits 
a pole at a certain $q$-value for $\omega =0$. 
In that case an excitation with arbitrary low energy 
is possible and the lattice becomes unstable with respect 
to a (static) deformation with wave number~$q$. 
From this condition the equation for the inverse
transition temperature $\beta_c=1/T_c$ easily follows as  
\begin{equation}\label{2-1TSP}
1=\kappa\int_0^\pi dk
\frac{\sin^2(k)\tanh[\frac{\beta_c}{2}\cos(k)]}{\cos(k)}\,,
\end{equation}
where $\kappa=g^2\omega_0/\pi$. A more detailed study of the complete
pole structure of  $\bar{D}^{\rm ret}(q,\omega)$ in both the
undimerized and dimerized phases is presented in Ref.~\cite{HFW00}, 
where also the ultrasound properties of the system were discussed. 
Here we focus on the temperature dependence of the 
dynamical structure factor above $T_c$,
\begin{figure}
\hfill\psfig{figure=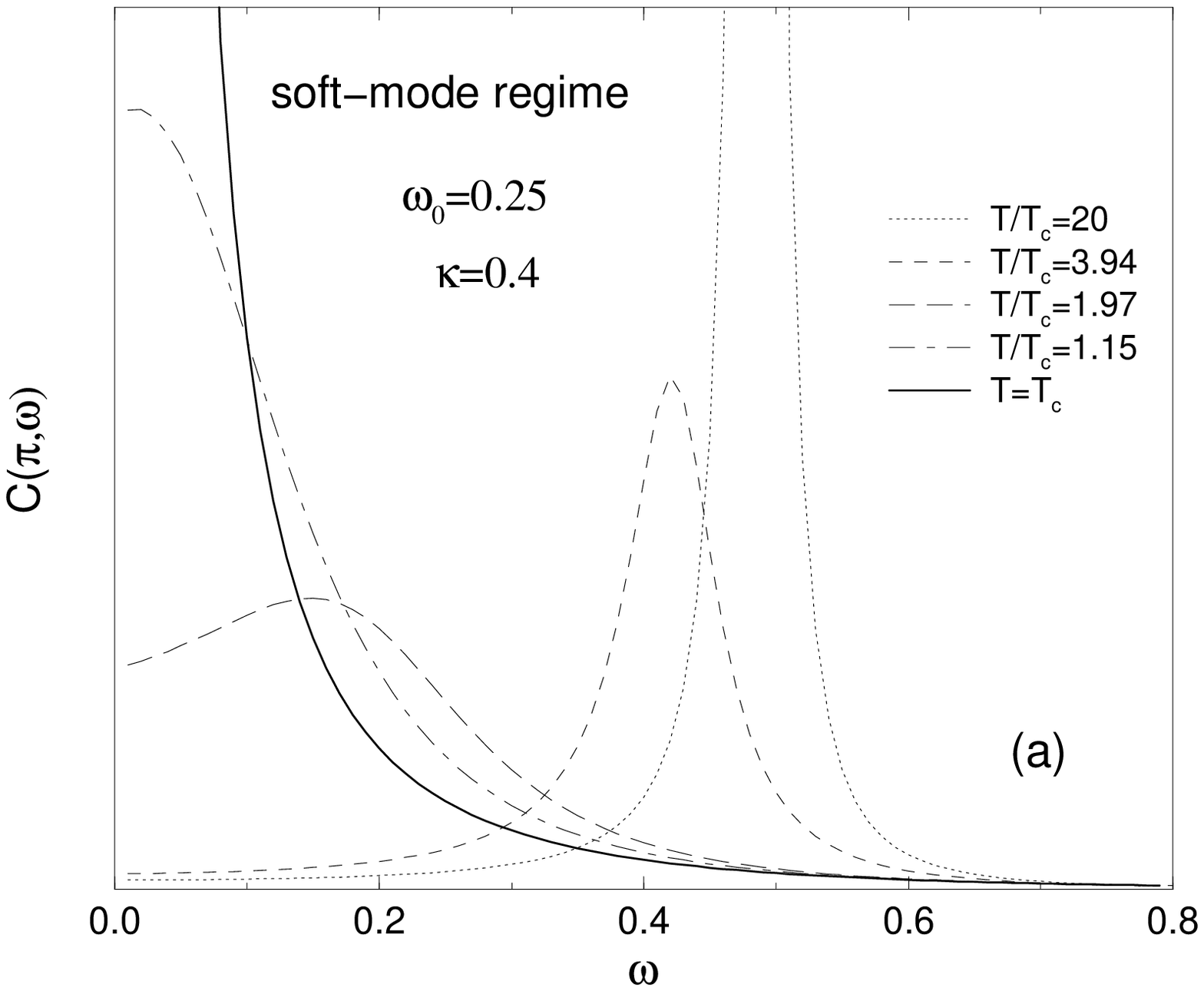,width=6.0cm}
\hfill\psfig{figure=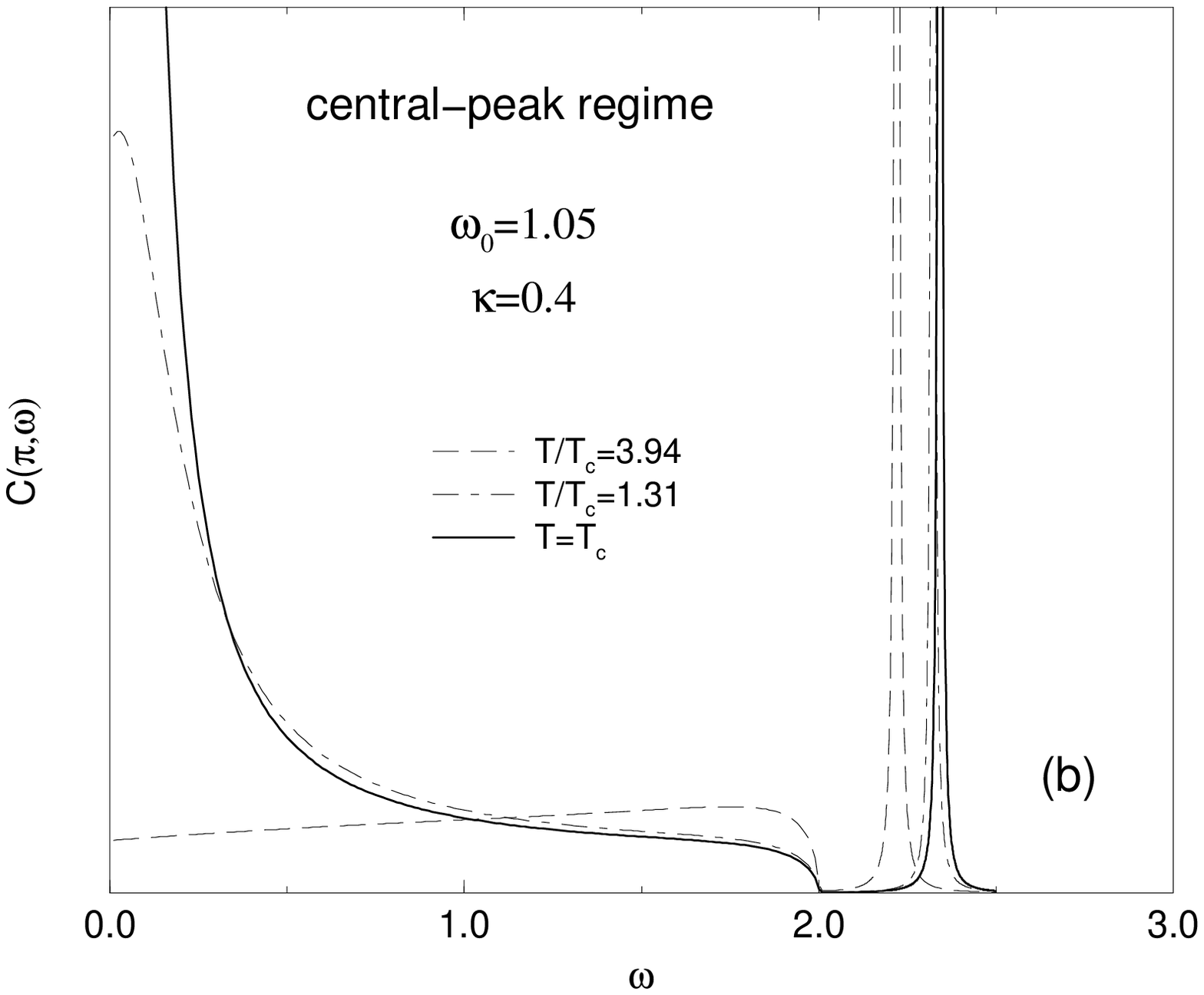,width=6.0cm}
\caption[fig5]{Dynamical structure factor for the magnetostrictive
XY model.}
\label{f5}
\end{figure}  
\begin{equation}
\label{A1struktfakt}
C(q,\omega)=-\frac{1}{\pi}\lim_{\delta\rightarrow 0}
\frac{\mathrm{Im}\ \bar{D}^{ret}(q,\omega+i\delta)}{1-e^{-\beta\omega}}\,,
\end{equation}
which is displayed in Fig.~\ref{f5} for two characteristic 
phonon frequencies $\omega_0=0.25$ and $\omega_0=1.05$, 
corresponding to the soft-mode (a) and central-peak (b) regimes, 
respectively. For small bare phonon frequencies the maximum in $C(\pi,\omega)$
is located around $\omega=\omega_{\pi}$ at high temperatures and moves
with decreasing temperatures to lower frequencies until a real divergence
of $C(\pi,\omega)$ appears for $T=T_c$ at $\omega = 0$.  
For large bare phonon frequencies a completely different behaviour is found.
Now the high-temperature phonon peak stays around $\omega=\omega_{\pi}$ and
even hardens in some degree as $T\to T_c$. At the same time 
a peak structure evolves at $q=0$ which becomes divergent at
$T=T_c$. This so-called central peak corresponds to a new 
collective magneto-elastic excitation of the coupled spin-phonon system,
which occurs at the SP transition and moves to higher energies as
the temperature is lowered further~\cite{HFW00}.
\section{Conclusions}
In this report the lattice dynamical effects on the Peierls
transition in coupled electron/spin-phonon systems were discussed.
Our approach was based on several generic model Hamiltonians obtained
from the one-dimensional Holstein/SSH Hubbard model in important 
limiting cases. Applying both basically exact numerical and 
approximative analytical methods we are able to calculate 
both ground-state and spectral properties of these simplified models
for the complete range of model parameters. The focus, however,
was on non-adiabatic effects due to the phonon dynamics.
The presented results confirm previous findings that quantum 
phonon fluctuations destroy the Peierls distorted state at 
sufficiently weak electron/spin-phonon interactions.

For the spinless fermion model this means that for weak electron-phonon
couplings the system resides in a metallic (gapless) phase described by two 
non-universal Luttinger liquid parameters. The renormalized 
charge velocity and the correlation exponent are obtained 
from finite-size scaling relations, fulfilled with great accuracy. 
The Luttinger liquid phase splits in an attractive and repulsive 
regime at low and high phonon frequencies, respectively. 
Here the polaronic metal, realized for repulsive interactions, 
is characterized by a strongly reduced mobility of the charge carriers.    
Increasing the electron-phonon coupling, the crossover 
between Luttinger liquid and charge-density-wave behaviour 
is found in excellent agreement with very recent DMRG results.
The transition to the CDW state is accompanied by significant changes
in the optical response of the system. Most notably seems 
to be the substantial spectral weight transfer from the 
Drude to the regular (incoherent) part of the optical conductivity, 
indicating the increasing importance of inelastic scattering processes 
in the CDW (Peierls distorted) regime.

As a simple model for a spin-Peierls system we considered  
a (frustrated) Heisenberg spin chain model supplemented by a
coupling to optical phonons with frequencies comparable 
to the magnetic exchange coupling, which is, 
e.g.. the relevant limit for the spin-Peierls
compound $\rm CuGeO_3$. The magnetic excitations inherently include
a local lattice distortion requiring a multi-phonon-mode treatment 
of the lattice degrees of freedom. When compared to the static
model of an alternating, dimerized spin chain the magnetic properties 
are strongly renormalized due to the coupled spin and lattice dynamics.
For the quantum spin-phonon model the dimerization dependence
of the spin triplet excitation gap is found to be in qualitative
agreement with experiment. In the anti-adiabatic phonon frequency range 
we used the concept of unitary transformations in order to integrate out 
phonon degrees of freedom, where the spin-phonon interaction 
results in an additional frustration in the effective spin-only model.  
For two types of spin-phonon couplings, the critical lines separating
gapless from gapped ground states are found to agree for 
the original spin-phonon and effective spin models.  
Finally the reliability of the RPA approach to SP transitions was
assessed in terms of an XY model with additional 
magneto-elastic interaction.
With increasing phonon frequency the dynamical displacement 
structure factor shows a crossover from a soft-mode to central-peak 
behaviour, normally linked to displacive and order-disorder types  
of transition respectively. The central peak predicted to appear at $T_c$
for large bare phonon frequency corresponds not to a real phononic but
rather to a new magneto-elastic excitation.  
\section*{Acknowledgements} 
The authors would like to thank B. B\"uchner, H. B\"uttner, 
R. J. Bursill,  A. P. Kampf, A. W. Sandvik, J. Schliemann, 
G. S. Uhrig, G. Wellein and R. Werner for valuable discussions.
The ED calculations were performed at the LRZ M\"unchen, 
NIC J\"ulich, and HLR Stuttgart. 

\end{document}